\documentstyle[twocolumn,epsfig,aps]{revtex}
\newcommand{\simle}
{\raisebox{-0.75ex}[-1.5ex]{$\;\stackrel{<}{\sim}\;$}}

\begin{document}

\draft
\title{
Theory of Anomalous Hall Effect in a Heavy fermion System \\
with a Strong Anisotropic Crystal Field}
\author{Hiroshi Kontani, Morio Miyazawa$^1$ and Kosaku Yamada$^1$}
\address{
Institute for Solid State Physics, University of Tokyo, 
7-22-1 Roppongi, Minato-ku, Tokyo 106}
\address{$^1$
Department of Physics, Faculty of Science, Kyoto University,
Sakyo-ku, Kyoto 606-01}

\date{\today}
\maketitle

\def\w{{\omega}}

\begin{abstract}
In a heavy fermion system, there exists the anomalous Hall effect
caused by localized $f$-orbital freedom, in addition to the 
normal Hall effect due to the Lorentz force.
In 1994, we found that the Hall coefficient caused by 
the anomalous Hall effect ($R_{\rm H}^{\rm AHE}$) is predominant
and the relation $R_{\rm H}^{\rm AHE} \propto \rho^2$ 
($\rho$ is the electrical resistivity) holds 
at low temperatures in many compounds.
In this work, we study the system
where the magnetic susceptibility is highly anisotropic
due to the strong crystalline electric field on $f$-orbitals.
Interestingly, we find that $R_{\rm H}^{\rm AHE}$
is nearly isotropic in general.
This tendency is frequently observed experimentally, 
which has casted suspicion that the anomalous Hall effect may be
irrelevant in real materials.
Our theory corresponds to corrections and
generalizations of the pioneering work on ferromagnetic metals
by Karplus and Luttinger.

\end{abstract}
%\pacs{PACS numbers:  75.10.Jm, 75.30-m, 75.40-s}

\vskip 0.6cm
\noindent
KEYWORDS \ : \
heavy fermion system, anomalous Hall effect, 
periodic Anderson model, orbital degeneracy, CeRu$_2$Si$_2$

%\narrowtext
\vskip 0.6cm
%%%%%%%%%%%%%%%%%%%%%%%%%%%%%%%%%%%%%%%%
% Definitions
%%%%%%%%%%%%%%%%%%%%%%%%%%%%%%%%%%%%%%%%

\def\s{{\sigma}}
\def\e{{\epsilon}}
\def\d{{\partial}}
\def\k{{ {\bf k} }}
\def\q{{ {\bf q} }}
\def\w{{\omega}}
\def\l{{\lambda}}
\def\L{{\Lambda}}       
\def\a{{\alpha}}
\def\b{{\beta}}
\def\g{{\gamma}}
\def\G{{\Gamma}}        
\def\v{{\varphi}}
\def\i{{ {\rm i} }}

%%%%%%%%%%%%%%%%%%%%%%%%%%%%%%%%%%%%%%%%
% INTRODUCTION
%%%%%%%%%%%%%%%%%%%%%%%%%%%%%%%%%%%%%%%%

In many heavy fermion (HF) systems, 
the Hall coefficient ($R_{\rm H}$)
in a paramagnetic state shows universal temperature dependence,
reflecting the crossover of the electronic property:
As temperature increases, $R_{\rm H}$ also increases rapidly,
and decreases after showing a peak at $T_0$.
 \cite{Onuki,Onuki2,Onuki3}
(we call $T_0$ the 'coherent temperature').
The peak value $R_{\rm H}(T_0)$ is apparently enhanced:
it reaches $\sim 10^2$ times that of a normal metal.
The origin of this interesting phenomena should be confirmed.

Today, some theoretical explanations have been proposed
in terms of the anomalous Hall effect (AHE),
which is caused by the $f$-orbital freedom.
 \cite{Kontani,Coleman,Fert}
In 1994, we proved the existence of the AHE 
in the periodic Anderson model (PAM) with orbital degeneracy,
using the Kubo formula and the Fermi liquid theory.
 \cite{Kontani,Kohno,Yamada}
The observed $R_{\rm H}$ in HF systems
is given by $R_{\rm H}= R_{\rm H}^{\rm AHE} + R_{\rm H}^{\rm norm}$,
where $R_{\rm H}^{\rm AHE}$ is the anomalous Hall coefficient we found
and $R_{\rm H}^{\rm norm}$ is the normal Hall coefficient
caused by the Lorentz force, which is temperature independent
below the coherent temperature.
 \cite{Fukuyama,Kohno2}
The AHE naturally explains the sign and the 
magnitude of $R_{\rm H}$ in HF systems,
and it shows a simple relation at low temperatures, 
$R_{\rm H}^{\rm AHE} = a \cdot \rho^2$, where
$\rho$ is the observed electrical resistivity.
(see Fig. 11 of ref. \cite{Kontani}.)
The mechanism of our AHE is essentially similar to that
found in ferromagnetic metal by Karplus and Luttinger.
 \cite{Kontani,Karplus}

\begin{figure}
 \epsfxsize=60mm \epsffile{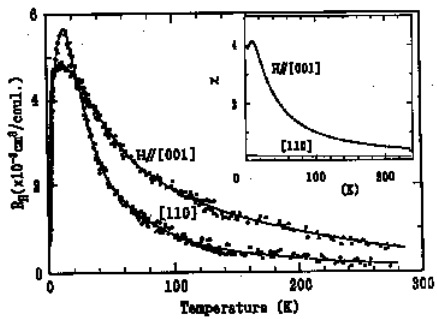}
%\vspace*{5.5cm}
\caption
{$R_{\rm H}$ vs $T$ for CeRu$_2$Si$_2$, borrowed from
ref. [3].
%ref.\cite{Onuki3}.
Here, $\rho$ is for $ [{\bar 1},1,0] $-direction.
We can see that $R_{\rm H}$ (also $\s_{ab}$) is nearly isotropic.
(Note that at $0.5$K $\rho=1.6\mu\Omega\cdot {\rm cm}$ for $a$-axis
and $0.81\mu\Omega\cdot {\rm cm}$ for $c$-axis.)
Inset: $\chi$ [$10^{-2}$eum/mol] vs $T$ [K] for CeRu$_2$Si$_2$. 
}
\label{fig:CeRuSi}
\end{figure}

\begin{figure}
 \epsfxsize=60mm \epsffile{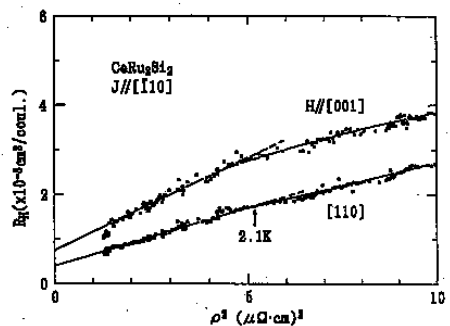}
%\vspace*{5.5cm}
\caption
{$R_{\rm H}$ vs $\rho^2$ for CeRu$_2$Si$_2$.
This figure is borrowed from 
ref. [3].
%ref.\cite{Onuki3}.
}
\label{fig:plot}
\end{figure}

In this work, we extend this theory to
an anisotropic system with a crystalline electric field (CEF).
Figure \ref{fig:CeRuSi} shows anisotropy of $R_{\rm H}$ and $\chi$
in CeRu$_2$Si$_2$
 \cite{Onuki3},
in which the relation $R_{\rm H} \propto \rho^2$ holds at low
temperatures, as shown by Fig. \ref{fig:plot}.
While $\chi$ is extremely anisotropic, $R_{H}$ is nearly isotropic.
This appears quite paradoxical because AHE is the magnetic contribution
and $R_{\rm H}^{\rm AHE}$ is enhanced by $\chi$ in the spherical system
\cite{Kontani}.
We calculate the direction-dependence of both 
$R_{\rm H}^{\rm AHE}$ and $\chi$ under the tetragonal CEF,
and find that $R_{\rm H}^{\rm AHE}$ is nearly
isotropic even if $\chi$ is highly anisotropic, in general.
This study confirms that AHE 
is predominant at low temperatures in many HF compounds.

The history of the study of AHE is very long,
which started on the ferromagnetic metals.
 \cite{Karplus,Smit,Kondo}.
Here, we comment on the other theory on AHE in 
paramagnetic HF materials given by Fert and Levy,
 \cite{Fert}
which is the improvement of ref.
 \cite{Coleman}.
They studied the impurity Anderson model ($J=5/2$)
with an additional scattering potential ($l=2$).
They calculated $R_{\rm H}^{\rm AHE}$ using the
Boltzmann equation for higher temperatures ($T > T_0$),
where each $f$-electron site is almost incoherent.
This mechanism is essentially the same as the theory of
AHE in dirty ferromagnetic metal, given by Smit.
 \cite{Smit}
They concluded that $R_{\rm H}^{\rm AHE} \propto \chi \cdot \rho$ 
for $T > T_0$.
Here we comment that their theory on AHE 
are essentially different from ours, which 
cannot be derived by the Boltzmann equation.
For example, the simple extrapolation of our theory 
predicts that $R_{\rm H}^{\rm AHE} \propto \chi$
in the incoherent region ($T>T_0$),
which is independent of $\rho$.
Both of them may co-exist.
But, it is definite that our AHE predominates at least for $T < T_0$.

\begin{figure}
%\vspace*{5cm}
%\epsfile{file=band.eps,height=5cm}
\epsfxsize=50mm \epsffile{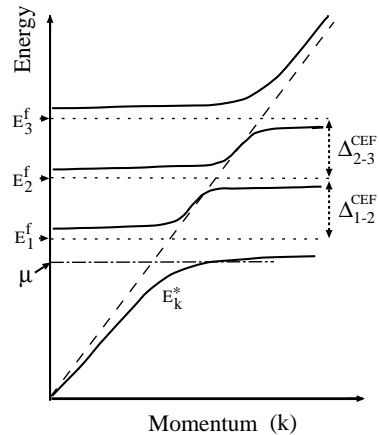}
\caption
{Effective band structure for $J=5/2$ PAM under the tetragonal CEF.
$\mu$ is the Fermi energy.
}
\label{fig:band}
\end{figure}

%%%%%%%%%%%%%%%%%%%%%%%%%%%%%
% Model and Theory
%%%%%%%%%%%%%%%%%%%%%%%%%%%%%
Here, we study Ce-compounds,
where $f^1$ configuration is realized.
Due to strong $l$-$s$ coupling,
% and strong Hund coupling,
we have only to treat $J=5/2$ sector, which has sixfold degeneracy.
In a tetragonal crystal, 
this degeneracy splits into three levels by CEF, 
$|\Gamma\rangle =|\pm1\rangle,|\pm2\rangle,|\pm3\rangle$, 
where $|+\G\rangle$ and $|- \G\rangle$ form a Kramers doublet.
( Now, $J_z$ is not a good quantum number.)
In this work, we discuss the orbitally degenerate ($J=5/2$) PAM
under a tetragonal CEF as follows:
\begin{eqnarray}
& & H = \sum_{\k \sigma} \epsilon_{\k} c_{\k \sigma}^\dagger c_{\k \sigma}
       + \sum_{\k \G\G'} E_{\G\G'}^f f_{\k \G}^\dagger f_{\k \G'}
       + \sum_{\k \G \sigma} ( V_{\k \G\sigma}^\ast f_{\k \G}^\dagger c_{\k \sigma}
 \nonumber \\
& &\ \ \ \ + {\rm h.c.} )
 + \frac U2 \sum_{\k\k'\q\G\G'}
 f_{\k+\q \G}^\dagger f_{\k'-\q \G'}^\dagger f_{\k' \G'} f_{\k \G},
 \label{eqn:hamiltonian} 
\end{eqnarray}
where $c_{\k \sigma}^\dagger$ ($f_{\k \G}^\dagger$) is the 
creation operator of the conduction electron ($f$-electron),
and $\e_\k$ ($E_{\G\G'}^f$) is the spectrum for conduction electrons
($f$-electrons).
In a tetragonal crystal,
the $f$-orbital eigenstate $|\G\rangle$ is given by
$|\G\rangle= \sum_M O_{\G M} |J\!=\!5/2,J_z\!=\!M\rangle$,
where $O_{\G M}$ is an $6\times6$ orthogonal matrix 
and is given by
\begin{eqnarray}
& &\ \ O_{\pm1,\pm5/2}= O_{\pm3,\pm3/2}= a, \nonumber \\
& &-O_{\pm1,\mp3/2}= O_{\pm3,\mp5/2}= \sqrt{1-a^2}, \label{eqn:CEF} \\
& &\ \ O_{\pm2,\pm1/2}=1, \nonumber
\end{eqnarray}
and other elements are zero.
In a cubic crystal, $a\!=\!\sqrt{5/6}\!=\!0.913$.
(In a hexagonal crystal, $a=1$.)
Then, we can easily verify that 
$V_{\k \G\sigma}= \sum_M O_{\G M} V_{\k M\sigma}$ and
$J_{\G\G'}^\g= \sum_{MM'} O_{\G M} O_{\G' M'} J_{MM'}^\g$ ($\g\!=\!x,y,z$),
$J^\g$ being the angular momentum along $\g$.
(Note that $V_{\k M\s}= V_0 \cdot \sqrt{4\pi/3} \sum_m -\s 
 \sqrt{ (7/2-M \s)/7 } \delta_{m,M-\s/2} 
 \cdot Y_{l=3}^m (\theta_k ,\varphi_k )$
is derived for a spherical case, where
$Y_{l=3}^m(\theta_k ,\varphi_k )$ is the spherical harmonic function.
\cite{Hanzawa})
Finally, $E_{\G\G'}^f$ under the magnetic field ${\bf H}$ is given by
$E_{\G\G'}^f= E_\G^f \cdot \delta_{\G\G'}
+ g \mu_B \sum_{\g} J_{\G\G'}^\g \cdot H^\g$
up to $H^1$-order,
where $g$ is the Lande's $g$-factor ( $g\!=\!6/7$ for $J\!=\!5/2$ ),
and we put $\mu_B=1$ hereafter.
For the time being, we limit our study to the case of $U\!=\!0$,
or to the mean-field (Gutzwiller) treatment:
$V_{\k\G\s} \rightarrow \sqrt{z_{\G}} \cdot V_{\k\G\s}$
and $E_\G^f \rightarrow {E_\G^f}^\ast \equiv z_{\G}\cdot(E_\G^f-\mu)+\mu$,
where $z_{\G}$ is the renormalization constant.
 \cite{Rice}
The electronic structure of orbitally degenerate $J=5/2$ PAM 
is shown in Fig. \ref{fig:band},
where $E_\k$ is the quasiparticle spectrum for $\k$,
given by the pole of Green functions.

% magnetic susceptibility
In the mean-field approximation, $\chi$ is given by
\begin{eqnarray}
& &\chi= \chi_{\rm P}+\chi_{\rm V}, 
  \label{eqn:susc} \\
& &\ \chi_{\rm P}= \sum_{\s}\sum_{\k}\left(\frac{\d E_{\k\s}}{\d H}
 \right)^2 \cdot \delta (\mu-E_{\k}), \nonumber \\
& &\ \chi_{\rm V}= -\sum_{\s}\sum_{\k} \frac{\d^2 E_{\k\s}}{\d H^2}
 \cdot \theta (\mu-E_{\k}), \nonumber
\end{eqnarray}
where $\chi_{\rm P}$ and $\chi_{\rm V}$ are called
Pauli susceptibility and Van-Vleck susceptibility, respectively.
Both $\chi_{\rm P}$ and $\chi_{\rm V}$ take the positive value
 \cite{Hanzawa}.
After a long calculation, we get the expressions
in the case of ${\bf H} \parallel \g$ ($\g=x,y,z$):
\begin{eqnarray}
& & \chi_{\rm P}^\g= g^2 \int \frac{{\rm d}\Omega_\k}{4\pi} \rho_\k^c(\mu)
 \sum_{\s\s'} a_\k(\mu) \cdot 
 \left| M_{\k\s\s'}^\g(\mu) \right|^2, 
  \label{eqn:susc-p} \\
& & \chi_{\rm V}^\g= -2 g^2 \int \frac{{\rm d}\Omega_\k}{4\pi}
 \int_{-D}^\mu \rho_\k^c(E_\k) dE_\k 
 \nonumber \\
& &\ \ \ \times \left\{ \ \frac{\d}{\d E_\k} 
 \left[ \frac12 a_\k(E_\k) \sum_{\s\s'}
 \left| M_{\k\s\s'}^\g(E_\k) \right|^2 \right] \right. 
 \nonumber \\
& &\ \ \ \ + \left. \sum_{\s,\G\G'\G''} \a_{\k\G\s}(E_\k) 
 \cdot \frac{J_{\G\G'}^\g J_{\G'\G''}^\g }{E_\k-E_{\G'}^f}
 \cdot \a_{\k\G''\s}^\ast(E_\k) 
 \ \right\},
  \label{eqn:susc-v}
\end{eqnarray}
where $\rho_\k^c(\e)$ represents the DOS for conduction electrons,
$-D$ is the bottom of $E_\k$, and 
\begin{eqnarray}
& &M_{\k\s\s'}^\g(\e)= \sum_{\G\G'} \a_{\k \G\s}(\e) 
 J_{\G\G'}^\g \a_{\k \G'\s'}^\ast(\e),          \\
& &\a_{\k \G \s}(\e)= V_{\k \G\s}/(\e-E_\G^f), \\
& &a_{\k}(\e)= \left( 1+ \sum_\G |\a_{\k \G \s}(\e)|^2 \right)^{-1}.
\end{eqnarray}
Note that $d E_\k / d \e_\k = a_{\k}(E_\k)$.
The $E_\k$-integration in (\ref{eqn:susc-v}) can be done easily
when $\rho_\k^c(\e)$ is constant.
In the case of no CEF ($E_\G^f=E^f$ for any $\G$), 
the relation $\chi_{\rm P} \approx \chi_{\rm V}$
is derived by (\ref{eqn:susc-p}) and (\ref{eqn:susc-v}).
In contrast, in the strong CEF case ($E_i^f \ll E_j^f,E_k^f$),
$\chi_{\rm P}$ shows strong anisotropy and $\chi_{\rm P} \gg \chi_{\rm V}$
is realized within the mean-field approximation.
In reality, it is not easy to investigate the 
accuracy of the above mean-field results.
In the case of no CEF,
this problem is studied by the Fermi liquid theory, 
 \cite{Kontani2}
or by the second-order perturbation theory w.r.t. U,
 \cite{Kontani3}
and the relation 
$\chi_{\rm P} \approx \chi_{\rm V}$ has been confirmed now.

% anomarous Hall conductivity
On the other hand, the Hall conductivity due to AHE
is given by eq. (2.16) in Ref. \cite{Kontani},
using the Kubo formula.
While we take the effect of CEF into account,
we neglect the vertex corrections for currents.
Under the magnetic field ${\bf H} \parallel \g$ ($\g\!=\!x,y,z$),
the anomalous Hall conductivity $\s_{\a\b}^{\rm AHE}$
($\e_{\a\b\g}\!=\!1$) is given by
\begin{eqnarray}
& &\s_{\a\b}^{\rm AHE}/H= g \frac {\rm e^2}{2} \sum_\s \sum_{\G\G'\G''}
 \int \frac{d\Omega_\k}{4\pi} \cdot \rho_\k^c (\mu) \cdot \frac1{\Delta_\k^c}
 \nonumber \\
& &\ \times \left[ \ v_{\k c}^\a \cdot 
% \a_{\k \G\s}^\ast(\mu+\i\Delta_\G) \a_{\k \G'\s}(\mu-\i\Delta_{\G'})
 \frac{V_{\k \G\s}^\ast V_{\k \G'\s}}
  {(\mu -E_\G^f +\i\Delta_\G)(\mu -E_{\G'}^f -\i\Delta_{\G'})}
 \right. \nonumber \\
& &\ \ \ \ \ \ \ \ \cdot
 \left( \frac{v_{\k \G'\G''}^\b J_{\G''\G}^\g}{\mu-E_{\G''}+\i\Delta_{\G''}}
 + \frac{J_{\G'\G''}^\g v_{\k\G''\G'}^\b}
  {\mu-E_{\G''}-\i\Delta_{\G''}} \right) 
 \nonumber \\
& &\ \ \ \ +  v_{\k c}^\b \cdot 
% \a_{\k \G\s}^\ast(\mu-\i\Delta_\G) \a_{\k \G'\s}(\mu+\i\Delta_{\G'})
 \frac{V_{\k \G\s}^\ast V_{\k \G'\s}}
   {(\mu -E_\G^f -\i\Delta_\G)(\mu -E_{\G'}^f +\i\Delta_{\G'})}
 \nonumber \\
& &\ \ \ \ \ \ \ \ \cdot \left.
 \left( \frac{v_{\k \G'\G''}^\a J_{\G''\G}^\g}{\mu-E_{\G''}-\i\Delta_{\G''}}
 + \frac{J_{\G'\G''}^\g v_{\k \G''\G'}^\a}{\mu-E_{\G''}+\i\Delta_{\G''}} \right)
 \ \right] \label{eqn:hallcond},
\end{eqnarray}
where
\begin{eqnarray}
& &v_{\k \G\G'}^\a
= \frac{\d}{\d k_\a} \left( \frac{\sum_\s V_{\k \G\s}^\ast
 V_{\k \G'\s}}{\mu-\e_\k} \right),
 \label{eqn:verocity} \\
& &v_{\k c}^\a = \sum_{\G\G'} \a_{\k\G\s}(\mu) 
 \a_{\k\G'\s}^\ast(\mu) \cdot v_{\k \G\G'}^\a.
\end{eqnarray}
The derivative of the numerator in (\ref{eqn:verocity}) 
gives the anomalous transverse current.
In (\ref{eqn:hallcond}), $\Delta_\G$ ($\Delta_\k^c$) is the 
damping rate of the $f$-electrons (conduction electrons), given by
\begin{eqnarray}
& &\Delta_\G= -{\rm Im}\Sigma_{\G}(\mu+\i0) 
 \label{eqn:damping} \\
& &\Delta_\k^c= \sum_{\G} |\a_{\k\G\s}(\mu)|^2 \cdot \Delta_\G,
\end{eqnarray}
where $\Sigma_{\G}(\w)$ is the self-energy
of $f$-electrons with $\G$, which is $\k$-independent 
within the local approximation. 
Here, $\Sigma_{\G}(\w)$ is brought by the Coulomb interaction $U$.
Within the second order perturbation and local approximation,
taking the Pauli principle for $\Gamma$ into account, 
it is easy to show that
\begin{eqnarray}
\Delta_\G = U^2 \cdot \frac\pi2 \rho_\G^f(\mu) \sum_{\G' \ne \G} 
 \left(\rho_{\G'}^f(\mu)\right)^2 \cdot (\pi T)^2
 \label{eqn:selfenergy}
\end{eqnarray}
for any $\G$ at low temperatures ($T\ll T_0$), where 
$\rho_\G^f(\mu)\equiv \sum_{\k \s} |\a_{\k\G\s}(\mu)|^2 \cdot \rho_\k^c(\mu)$ 
is the density of states for $f$-electrons with $\G$ at the Fermi energy.
The relation (\ref{eqn:selfenergy}) tells that
$\Delta_\G/\Delta_{\G'}$ is temperature-independent 
for any $\G$ and $\G'$ at low temperatures,
which will also be expected in the coherent region ($T\simle T_0$).
As a result, $\s_{\a\b}^{\rm AHE}$ given by (\ref{eqn:hallcond})
is temperature-independent in the coherent region,
where $\chi$ remains nearly constant and
$\Delta_{\G} \ll (E_\G^f-\mu)$ is realized.
 \cite{Kontani}
The definition of the Hall coefficient for ${\bf H} \parallel \g$ is
$R_{\rm H}= (\s_{\a\b}/H)/(\s_{\a\a}\s_{\b\b})$
(where $\e_{\a\b\g}\!=\!1$).
Thus, the relation $R_{\rm H}^{\rm AHE} \propto \rho^2$ is confirmed,
as shown in Fig. \ref{fig:plot}.

The theory of Karplus and Luttinger contains a serious inconsistency,
pointed out by ref. 
\cite{Smit}.
In ref. \cite{Kontani},
we solved it by taking the dissipation of quasiparticles into account.
The new aspects of the AHE we find are as follows:
(i) We derived the exact form of $\s_{\a\b}^{AHE}$, and found that 
it depends on the ratio $\Delta_\G/\Delta_{\G'}$, which
differs greatly from $1$ in the anisotropic system.
This means that $\s_{\a\b}^{AHE}$ depends on the origin of 
the quasi-particle damping.
(ii) In our theory, the orbital moment ($g{\hat J}^\g$)
plays an important role, and the $l$-$s$ coupling term is not essential.

\begin{figure}
%\epsfile{file=fig1.eps,height=6cm}
\epsfxsize=60mm \epsffile{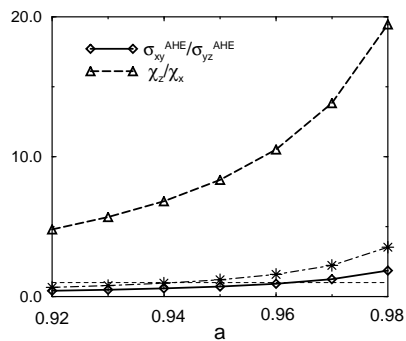}
%\vspace*{5cm}
\caption
{$a$-dependence of the anisotropy of $\chi$ and $\s^{\rm AHE}$.
($a$ characterizes the CEF introduced by 
eq. (2).)
%(\ref{eqn:CEF}).
Here, we set $\Delta_{1\mbox{-}2}=0.3$ and $|E_1^f-\mu|=0.03$.
In comparison, we also show $\s_{yz}^{\rm AHE}/\s_{xy}^{\rm AHE}$ 
for the case of $\Delta_\G={\rm constant}$, instead of 
eq. (14),
%(\ref{eqn:damping})
by stars (dash-dotted line).
}
\label{fig:aniso1}
\end{figure}

\begin{figure}
%\epsfile{file=fig2.eps,height=6cm}
\epsfxsize=60mm \epsffile{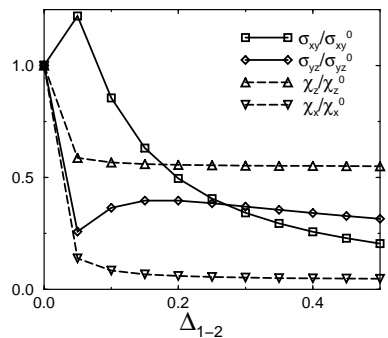}
%\vspace*{5cm}
\caption
{$\Delta_{\rm CEF}$-dependence of $\chi^z$, $\chi^x$,
$\s_{xy}^{\rm AHE}$ (${\bf H} \parallel {\bf z}$) and 
$\s_{yz}^{\rm AHE}$ (${\bf H} \parallel {\bf x}$).
Here, we set $a=0.96$
and $|E_1^f-\mu|=0.03$.
}
\label{fig:aniso2}
\end{figure}

% numerical calcurations
Following are the results of our numerical calculations
on $\chi$ and $\s_{\a\b}^{AHE}/H$, given respectively by 
(\ref{eqn:susc}) and (\ref{eqn:hallcond}).
%Unfortunately, we do not know sufficient physical parameters of
%CeRu$_2$Si$_2$.
Now we put $z_{\G}=1$.
Here, we study the case $\Delta_{1\mbox{-}2},\Delta_{2\mbox{-}3}>0$, 
where $\Delta_{1\mbox{-}2} = E_2^f-E_1^f$ and 
$\Delta_{2\mbox{-}3} = E_3^f-E_2^f$.
Taking the experimental data on CeRu$_2$Si$_2$ into account,
we put 
$\Delta_{2\mbox{-}3}= 3 \cdot\Delta_{1\mbox{-}2}$
%$\Delta_{2\mbox{-}3}= 5 \cdot\Delta_{1\mbox{-}2}$
 \cite{CEF}.
For simplicity, we treat $\Delta_{1\mbox{-}2}$ and the parameter $a$
independently.
Other parameters we use are
$E_1^f \!=\! -0.2$, $\mu \!=\! -0.23$, and $V_0 \!=\! 0.2$.
We also put $\e_\k= -1+2\cdot(\k/\pi)^3$, where
$\rho_k^c = k^2 \cdot d\e_k/dk$ is $\k$-independent.
These selections (and the shape of the Fermi surface)
is to be irrelevant to the conclusion qualitatively.
Figure \ref{fig:aniso1} shows the $a$-dependence of 
$\chi^z/\chi^x$ and $\s_{xy}^{\rm AHE}/\s_{yz}^{\rm AHE}$ 
in the case of $\Delta_{1\mbox{-}2}=0.3$.
We can recognize the isotropic feature of the anomalous Hall coefficient,
which is quantitatively affected by the form of the damping rate.
(When $\Delta_{1\mbox{-}2}=\Delta_{2\mbox{-}3}=0$, 
both $\chi$ and $\s_{ab}$ are isotropic and independent of $a$.)
Figure \ref{fig:aniso2}
shows the dependence of $\chi^z$, $\chi^x$, 
$\s_{xy}^{\rm AHE}$ and $\s_{yz}^{\rm AHE}$ on $\Delta_{1\mbox{-}2}$,
in the case of $a=0.96$.
We can see that $R_{\rm H}$ is less affected by
$\Delta_{1\mbox{-}2}$ than $\chi$.
As a result, $R_{\rm H}^{\rm AHE}$ can take a large value
even in strongly anisotropic systems.
Especially, $R_{\rm H}^{\rm AHE}$ tends to be isotropic 
for larger $\Delta_{1\mbox{-}2}$ ($\Delta_{1\mbox{-}2} \gg |E_1^f-\mu|$).
%In Fig.\ref{fig:aniso1} and \ref{fig:aniso2},
%we can see that results depend strongly on the form of damping rare.

Such a difference of the anisotropy between on $\chi$
and on $\s_{\a\b}^{AHE}$ can be understood naturally as follows:
$\chi_{\rm P}$, given by (\ref{eqn:susc-p}), comes mainly from 
the lowest Kramers doublet (say, $\G=\pm1$).
On the contrary, higher Kramers doublets are indispensable
for $\s_{\a\b}^{AHE}$, given by (\ref{eqn:hallcond}),
because AHE is the multi-band contribution.
 \cite{Kontani}
Thus, $\s_{\a\b}^{AHE}$ is averaged and is nearly independent of 
directions.
In conclusion, we find that the anisotropies of $\chi$ and 
$R_{\rm H}^{\rm AHE}$ can be quite different.
In the strongly anisotropic system due to the (tetragonal) CEF,
in general, $\chi$ is strongly anisotropic and 
$R_{\rm H}^{\rm AHE}$ tends to be isotropic.

% Future Problems
Finally, we point out some remaining problems.
The present model may be too simplified for quantitative study
in that it contains only one spherical conduction band $\e_\k$.
Besides, some kinds of vertex corrections (or $1/z_\G$) 
omitted in this work should be studied for complete analysis.
Moreover, the consideration on the Umklapp processes will be 
necessary for some compounds:
References \cite{Kontani,Yamada} show that 
$\s_{xx} \propto \Delta_c^{-1} \cdot C_{\rm UM}$ and
$\s_{xy}^{\rm AHE}/H \propto \{C_{\rm UM}\}^2$, thus
$R_{\rm H} \propto \Delta_c^2 \cdot \{C_{\rm UM}\}^0$,
where the coefficient $C_{\rm UM}$ reflects the Umklapp processes
for electron-electron scattering,
which dominate the various transport phenomena at low temperatures.
 \cite{Yamada2}
(Note that the Umklapp processes are irrelevant to $R_{\rm H}$.)
$C_{\rm UM} \sim O(1)$ is satisfied in the usual case.
But $C_{\rm UM} \sim 10$ will be realized in CeRu$_2$Si$_2$,
which shows large $\s_{xy}^{\rm AHE}/H$ at low temperatures.
(see Fig. \ref{fig:plot} or ref. \cite{Kontani}.)
%Thus, in order to understand the observed value of 
%$\s_{\a\b}^{AHE}$ or $\s_{\a\a}$ further,
%we have to study on the Umklapp processes (which is rather 
%irrelevant to $R_{\rm H}$).
At last, we comment on the existence of the AHE in other kinds of metals.
For example, at low temperatures, $R_{\rm H}$ in Sr$_2$RuO$_4$ shows 
the temperature dependence similar to that in HF compounds.
 \cite{SrRuO}
We find that the relation $R_{\rm H} \propto \rho^2$ 
holds for $T\simle6$K.
 \cite{comment}
%In Sr$_2$RuO$_4$, $|R_{\rm H}(6{\rm K})-R_{\rm H}(0{\rm K})|$ is small
%because the enhancement of $\chi$ is also small.
We believe that the AHE discussed in this work 
is general to many non-magnetic metals with orbital-freedom.

% Acknowledgements
We would like to thank Kazuo Ueda for valuable comments.
We also thank to Yoshihito Miyako and Y. Tabata for experimental 
data on CeRu$_2$Si$_2$.
This work is financially supported by a Grant-in-Aid for Scientific
Research on Priority Areas from the Ministry of Education, 
Science, Sports and Culture.

%%%%%%%%%%%%%%%%%%%%%%%%%%%%%%%%%%%%%%%%%%%%%%%%%%%%%%%%%%%%%%%%%%%%%%%%

%%%%%%%%%%%%%%%%%%%%%%%%%%%%%%%%%%%%%%%%%%%%%%%%%%%%%%%%%%%%%%%%%%%%%%%%

% FIG1

% FIG2

% FIG3

%%%%%%%%%%%%%%%%%%%%%%%%%%%%%%%%%%%%%%%%%%%
% FIG4 : ~/ginnan/AHE97/figure2/fig1.eps
%%%%%%%%%%%%%%%%%%%%%%%%%%%%%%%%%%%%%%%%%%%

%%%%%%%%%%%%%%%%%%%%%%%%%%%%%%%%%%%%%%%%%%%%%%
% FIG5 : ~/ginnan/AHE97/figure2/fig1.eps
%%%%%%%%%%%%%%%%%%%%%%%%%%%%%%%%%%%%%%%%%%%%%%


\begin{references}
\bibitem{Onuki}
 Y. {\= O}nuki, T. Yamayoshi, I. Ukon, T. Komatsubara, A. Umezawa, W. K. Kwok,
 G. W. Crabtree and D. G. Hinks:
 J. Phys. Soc. Jpn. \boldmath {\bf 58} (1989) 2119.
\bibitem{Onuki2}
 Y. {\= O}nuki, T. Yamayoshi, T. Omi, I. Ukon, A. Kobori and T. Komatsubara:
 J. Phys. Soc. Jpn. {\bf 58} (1989) 2126.
\bibitem{Onuki3}
 Y. {\= O}nuki, S.W. Yun, K. Satoh, H. Sugawara and H. Sato:
in {\it Transport and Thermal Property of $f$-electron Systems},
ed. Oomi et.al. (Plemum Press (1993) p103).
\bibitem{Kontani}
 H. Kontani and K. Yamada: J. Phys. Soc. Jpn. {\bf 63} (1994) 2627.
\bibitem{Coleman}
 P. Coleman, P. W. Anderson and T. V. Ramakrishman:
 Phy. Rev. Lett. {\bf 55} (1985), 414.
\bibitem{Fert}
 A. Fert and P. M. Levy: Phy. Rev. B. {\bf 36} (1987) 1907.
\bibitem{Kohno}
 H. Kohno and K. Yamada:
 J. Magn. \&Magn. Matter. {\bf 90} \& {\bf 91} (1990) 431.
\bibitem{Yamada}
 K. Yamada, H. Kontani, H. Kohno and S. Inagaki:
 Prog. Theor. Phys. {\bf 89} (1993), 1155.
%\bibitem{Kasuya}
% T. Kasuya, J. Phys. Soc. Jpn.  ....
\bibitem{Fukuyama}
 H. Fukuyama, H. Ebisawa and Y. Wada: Prog. Theor. Phys. {\bf 42} (1969) 494.
\bibitem{Kohno2}
 H. Kohno and K. Yamada: Prog. Theor. Phys. {\bf 80} (1988) 623.
\bibitem{Karplus} 
 R. Karplus and J. M. Luttinger: Phys. Rev. {\bf 95} (1954) 1154.
\bibitem{Smit}
 J. Smit: Physica {\bf 21} (1955) 877 ; {\bf 24} (1958) 39.
\bibitem{Kondo}
 J. Kondo: Prog. Theor. Phys. {\bf 27} (1962) 772.
%\bibitem{Hanzawa2}
% K. Hanzawa, Y. Yosida and K. Yamada, Prog. Theor. Phys. {\bf 81} (1989) 960.
\bibitem{Hanzawa}
 K. Hanzawa, Y. Yosida and K. Yamada: Prog. Theor. Phys. {\bf 77} (1987) 1116,
 and references are therein.
\bibitem{Rice}
 T.M. Rice and K. Ueda: Phys. Rev. Lett. {\bf 55} (1985) 995, 2093(E).
\bibitem{Kontani2} 
 H. Kontani and K. Yamada: J. Phys. Soc. Jpn. {\bf 65} (1996) 172.
\bibitem{Kontani3}
 H. Kontani and K. Yamada: J. Phys. Soc. Jpn. {\bf 66} (1997) No.8.
\bibitem{Yamada2} 
 K. Yamada and K. Yosida: Prog. Theor. Phys. {\bf 76} (1986) 621:
$C_{\rm UM}$ is determined by the shape of the Fermi surface and 
the Brillouin zone, and it shows direction-dependence in reality.
\bibitem{SrRuO}
 A.P. Mackenzie, N.E. Hussey, A.J. Diver, S.R. Julian, Y. Maeno,
 S. Nishizaki and T. Fujita: Phy. Rev. B. {\bf 54} (1996) 7425.
\bibitem{comment}
 present authors (unpublished).
\bibitem{CEF}
 Y. Tabata and Y. Miyako, they expect 
 $\Delta_{1\mbox{-}2} \sim$ 245K and $\Delta_{2\mbox{-}3} \sim$ 755K
 in CeRu$_2$Si$_2$, from the observed $\chi$ at higher temperatures
 (unpublished).
\end{references}
\end{document}